\documentclass[structabstract]{aa}  
\usepackage{graphicx}
\usepackage{color}
\usepackage{epsfig}
\usepackage{longtable,lscape}
\usepackage[varg]{txfonts}
\usepackage{natbib}
\bibpunct{(}{)}{;}{a}{}{,}  
\usepackage{rotating}

\def\kms {\rm{km~s^{-1}}}
\def\apj {ApJ}
\def\apjl {ApJL}
\def\apjs {ApJS}
\def\aj {AJ}
\def\aap {A\&A}
\def\mnras {MNRAS}

\def\Mvir {\mathcal{M}_{vir}}

\begin{document}
\title{Comparing galaxy populations in compact and loose groups of galaxies III. 
Effects of environment on star formation}
\author{Valeria Coenda, Hern\'an Muriel and H\'ector J. Mart\'inez}
\institute{Instituto de Astronom\'ia Te\'orica y Experimental (IATE), CONICET, CCT C\'ordoba,
Laprida 854, C\'ordoba, X5000BGR. Argentina.\\
Observatorio Astron\'omico, Universidad Nacional de C\'ordoba,
Laprida 854, C\'ordoba, X5000BGR. Argentina.\\
\email{vcoenda@oac.unc.edu.ar}}
\date{Received XXX, XXXX; accepted XXX , XXXX }
\abstract
   {}
{This paper is part of a series in which we perform a systematic comparison of the galaxy properties
inhabiting compact groups, loose groups and the field. In this paper we focus our study to the age 
and the star formation in galaxies.}
{For galaxies in selected samples of compact groups, loose groups and field, we compare the 
distributions of the following parameters: D$_n(4000)$ as an age indicator, and the specific star 
formation rate as indicator of ongoing star formation. We analyse the dependence of these
parameters on galaxy type, stellar mass and, for group galaxies, their dependence on the dynamic 
state of the system. We also analyse the fraction of old, and of high star forming galaxies as a 
function of galaxy stellar mass in the environments we probe.}
{Galaxies in compact groups have, on average, older stellar populations than their loose group or 
field counterparts. Early-type galaxies in compact groups formed their stars and depleted their gas 
content more rapidly than in the other environments. We have found evidence of two populations of 
late-type galaxies in dynamically old compact groups: one with normal specific star formation rates 
and another with markedly reduced star formation.}
{Processes that transform galaxies from star forming to quiescent act upon galaxies faster and more 
effectively in compact groups. The unique characteristics of compact groups make them an extreme 
environment for galaxies, where the transition to quiescence occurs rapidly.}
\authorrunning{Coenda, Muriel \& Mart\'inez}
\titlerunning{Environment \& star formation }
\maketitle
\section{Introduction} 
\label{sec:intro}
One of the major challenges of the modern astrophysics is to determine when and where stars form.
The well known morphology-density relation (\citealt{Oemler:1974}, \citealt{Dressler:1980}) clearly 
indicates that the environment plays a fundamental role in the evolution of galaxies. Moreover, it 
has been suggested  by \citet{Kauffmann:2003} that the galaxy property most sensitive to environment
is the star formation rate by unit mass (specific star formation rate, SSFR). 
These authors found that galaxies with low stellar masses, show a decrease of the SSFR by more than 
a factor of 10 from low to high density. This decrease is less marked for high-mass galaxies. Models
of galaxy formation tend to reproduce the dependence of the star formation rate, SFR, with the local
density, nevertheless, they fail to reproduce the fraction of quiescent galaxies. 
\citet{Hirschmann:2014} found that models slightly under-estimate the quiescent fraction of central 
galaxies and significantly over-estimate that of satellite galaxies. 

Although the local density can be a suitable way of representing the environment, similar values of
local density can be found in regions with different physical conditions for both, the dynamics of 
galaxies and the intergalactic medium. A good example of this situation are compact groups (CGs) and
the core of rich clusters of galaxies. These two environments may have similar local density but 
very different velocity dispersion and gas temperature. On the other hand, similar velocity 
dispersion can be found in environments with very different local density, as is the case of CGs and
loose groups (LGs). Therefore, the comparative study of the galaxy properties in different types of 
systems of galaxies has become a powerful tool to understand the effects of the environment in 
galaxy evolution. These analyses have been benefited by the large-scale galaxy surveys, such as the
Sloan Digital Sky Survey (SDSS, \citealt{York:2000}) that have allowed the identification of 
thousands of groups, CGs and clusters of galaxies.

In \citet{Coenda:2012} (hereafter paper I) and \citet{Martinez:2013} (hereafter paper II) we 
performed a detailed comparison between the properties of galaxies in compact and loose groups. In 
paper I, we found that the fraction of red and early-type galaxies is larger in CGs. Our results 
suggest that galaxies in CGs are, on average, systematically smaller in size, more concentrated, 
and have higher surface brightness than galaxies in the field or in LGs. In Paper II we compared the
properties of the brightest group galaxy (BCGs) and found that BCGs in CGs are brighter, more 
massive, larger, redder and more frequently classified as elliptical. We concluded that galaxies 
inhabiting CGs have undergone a major transformation compared to galaxies that inhabit LGs. 

Traces of the different paths in the evolution of galaxies can be found in terms of differences in 
the ages of the stellar populations. \citet{Proctor:2004} and \citet{MdO:2005} compared the ages of
galaxies in CGs and in other dense environments and found that galaxies in CGs tend to be older. The
high density and low velocity dispersion in the CG environment set up the ideal conditions for 
gravitational interactions and, eventually, a favourable scenario for the truncation of the star 
formation. \citet{delaRosa:2007} found evidences of star formation quenching in low mass galaxies in
CGs. These authors confirmed that galaxies in CGs are older than in the field. Based on near and 
mid-infrared imaging, \citet{Bitsakis:2010} studied the SSFR in CGs, isolated galaxies and 
interacting pairs and found no differences. However, for dynamically old systems, they found that 
late-type galaxies show slightly lower SSFR than in dynamically young CGs. They attributed this 
effect to the multiple past interactions among the galaxies in dynamically old groups. Similar 
results were found by \cite{Bitsakis:2011} performing a multiwavelengh analysis from the ultraviolet
to the infrared. Other peculiarities have been reported for galaxies in CGs. \citet{Tzanavaris:2010}
found a significant bimodality in the SSFR with a gap between low and high-SSFR (see also 
\citealt{Walker:2012}). These results are interpreted as further evidence that the high galaxy 
densities and low galaxy velocity dispersion play a important role in accelerating galaxy evolution
by enhancing star formation and bringing on a fast transition to quiescence. A bimodal behaviour in
SSFR has also been reported in LGs and clusters (\citealt{Wetzel:2011}). 
\begin{figure}[t]
\centering
\includegraphics[width=9cm]{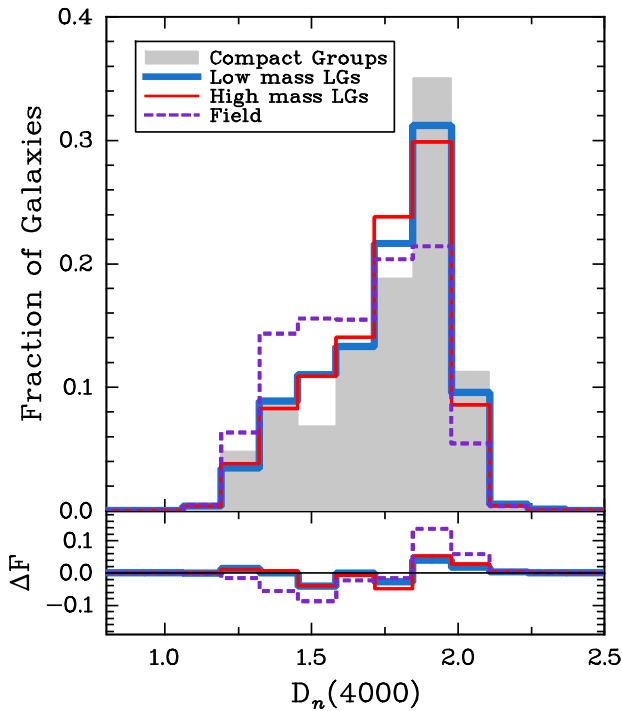}
\caption{Distributions of D$_n(4000)$ parameter in our samples. CGs ({\em grey shaded histograms}), 
the field ({\em violet dashed line}), low-mass LGs ({\em thick blue solid line}) and high-mass LGs 
({\em red solid line}). All distributions have been normalised to have the same area. In the panel 
below, we show the residuals between the distribution corresponding to galaxies in CGs and in the
other environments. The binsize is 0.13.} 
\label{fig:d4000_tot}
\end{figure}
The stellar populations of galaxies in CGs has also been studied using spectral synthesis analyses. 
\cite{Plauchu:2012} found that early-type galaxies show a higher content of old and intermediate 
stellar populations. They also reported that early-type spirals in CGs show lower SSFR values, while
late-type spirals peak at higher values compared with their counterparts in isolation and support 
the scenario where galaxies in CGs have evolved more rapidly than isolated galaxies.

It has been suggested that the large scale structure can also affect the evolution of galaxies. 
\citet{Scudder:2012} compared the galaxy properties in a sample of CGs that they further split by 
the large scale environment into isolated and embedded subsamples. They found that the SFR of star 
forming galaxies in CGs are significantly different between isolated and embedded systems, being 
higher in the first environment.

The galaxy properties in LGs have been extensively studied and it has been shown that the group 
environment is also efficient in quenching the star formation (e.g. \citealt{MZDML:2002}; 
\citealt{Bai:2010}; \citealt{Wetzel:2011}; \citealt{McGee:2011}; \citealt{Rasmussen:2012}; 
\citealt{Wetzel:2013}). \citealt{Wetzel:2013} study the quenching timescale of satellite galaxies 
and found that the SFR in groups evolve unaffected for 2-4 Gyrs after the infall, and then, the star
formation quenches rapidly. Similar timescales have been estimated by \citet{McGee:2011}. 

It becomes clear that both loose and compact groups are efficient in quenching the star formation in
galaxies, nevertheless it is not yet clear how the compactness differences moderate or increase the 
quenching process. As it was done in Papers I and II we propose a systematic comparison between the
properties of galaxies in compact and loose groups and field galaxies. In this paper we focus on the
star formation rate and the age of galaxies. For this purpose we use the MPA-JHU DR7 release of 
spectra measurements (http://www.mpa-garching.mpg.de/SDSS/DR7/). Using line fluxes, continuum 
indexes, line widths, etc. the MPA-JHU catalogue provides stellar masses (\citealt{Kauffmann:2003}),
star formation rate (\citealt{Brinchmann:2004}) and gas-phase metallicity (\citealt{Tremonti:2004}) 
for the the Main Galaxy Sample (MGS; \citealt{Strauss:2002}) of the Seventh Data Release of the SDSS
(DR7, \citealt{dr7}).

This paper is organised as follows: in section \ref{sec:sample} we describe the samples of groups 
and galaxies used;  in section \ref{sec:comparison} we perform comparative studies of the galaxy 
populations in CGs, LGs and the field; finally, our results are summarised and discussed in section 
\ref{sec:discussion}. Throughout the paper, a flat cosmological model is assumed, with parameters 
$\Omega_0=0.3$, $\Omega_{\Lambda}=0.7$, and a Hubble's constant $H_0=100~h~\kms~{\rm Mpc}^{-1}$. 
All magnitudes were corrected for Galactic extinction using the maps by \citet{sch98} and are in 
the AB system. Absolute magnitudes and galaxy colours were $K-$corrected using the method of 
\citet{Blanton:2003}~({\small KCORRECT} version 4.1).
\begin{figure}[t]
\centering
\includegraphics[width=9cm]{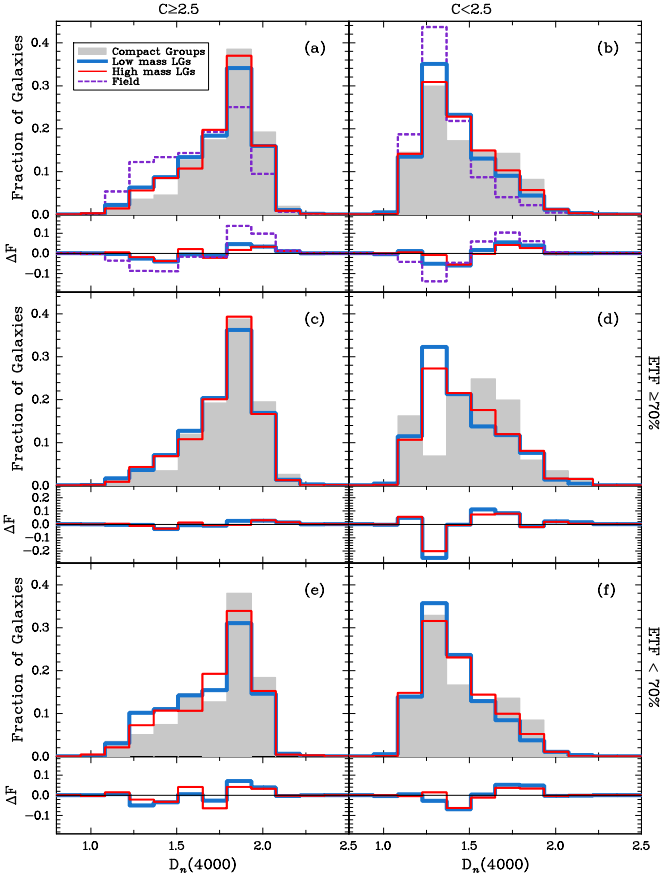}
\caption{Distributions of D$_n(4000)$ parameter in our samples: {\em Left panels} include early-type
galaxies according to their concentration parameter, while {\em right panels} include late-type 
galaxies in our samples. {\em Panels} (a) and (b) show all galaxies in the samples, {\em panels} 
(c) and (d) galaxies in groups with ETF$\ge70$\%, while {\em panels} (f) and (g) galaxies in groups 
with ETF$<70$\%. All distributions have been normalised to have the same area. Below each panel we 
show the residuals between the CGs and the other distributions. The binsize is 0.14.
Based on KS tests, we find there is no pair of them drawn from the same underlying distribution. Line types as in Fig. \ref{fig:d4000_tot}.} 
\label{fig:d4000}
\end{figure}
\section{The samples}
\label{sec:sample}

\subsection{The sample of compact and loose groups}
\label{subsec:grsample}
The samples of compact and loose groups of galaxies used in this paper have been constructed
from the same group catalogues used in Paper I and following the same selection criteria.

The sample of compact groups is drawn from catalogue A of 
\citet{McConnachie:2009}, who used the original selection criteria of \citet{Hickson:1982} to 
identify compact groups in the sixth data release of the SDSS \citep{dr6}.The catalogue A has 2,297
groups, adding up to 9,713 galaxies, down to a Petrosian (\citealt{petro76}) limiting magnitude of 
$r=18$, and has spectroscopic information for 4,131 galaxies (43\% completeness). In this work, we
select all groups in the A catalogue within the redshift range $0.06\le z\le0.18$, which have 
spectroscopic redshift for at least one member galaxy that have a counterpart in MPA-JHU data set. 
This particular choice of redshifts was defined in Paper I aiming at a fair comparison
with two samples of loose groups, of low and high mass, which we describe below.  
The number of low mass groups at higher redshifts rapidly drops and, conversely does the
number of high mass groups at lower redshifts.
We also restrict our analyses to galaxy members with apparent magnitudes $14.5\le r\le17.77$, 
i.e., the range in which the Main Galaxy Sample (MGS; \citealt{Strauss:2002}) is complete. This 
results in a sample of 778 compact groups adding up to 978 galaxies.

The samples of loose groups are drawn from the sample of groups identified by 
\citet{ZM11} (hereafter ZM11) in the MGS in SDSS DR7. They used a friends-of-friends algorithm 
\citep{H&G:1982} to link galaxies into groups, followed by a second identification using a higher 
density contrast in groups with at least 10 members, in order to split merged systems and clean up 
any spurious member detection. The authors computed group virial masses from the virial radius of 
the systems and the velocity dispersion of member galaxies \citep{Limber:1960,Beers:1990}. The 
sample of ZM11 comprises 15961 groups with more than 4 members, adding up to 103342 galaxies. As 
in Paper I, we split the sample of loose groups into two subsamples of low, 
$\log(\Mvir/M_{\odot}h^{-1})\le 13.2$, and high, $\log(\Mvir/M_{\odot}h^{-1})\ge 13.6$ virial mass. 
This choice excludes about 30\% of the groups in ZM11, those of intermediate virial mass,
that is within 0.2 dex of the median virial mass of the ZM11 catalogue at 
$\log(\Mvir/M_{\odot}h^{-1})\sim 13.4$.
Analogously to Paper I, in order to perform a fair comparison between the galaxies of CGs and LGs, 
we use a Monte Carlo algorithm to randomly select groups from these two subsamples. These new 
subsamples of low and high mass LGs have similar redshift distributions to that of the CGs. Our 
final subsamples of low and high mass LGs include 2319 and 2352 systems, adding up to 6996 and 
7055 galaxies, respectively.

The samples of groups and galaxies used in this study are not exactly the same  
as in Paper I because here we restrict our analyses to galaxies which have data from the
MPA-JHU DR7. This means that not all the galaxies we used in Paper I are included now, which in 
turn, implies that about 9\% of the compact groups in Paper I are now ruled out since they have
not a single galaxy with  MPA-JHU DR7 data. The exclusion of these compact groups determines
minor changes in the loose group samples which are bound to have redshift distributions 
similar to that of the compact groups.

We refer the reader to Paper I for further details.
\begin{figure*}[th]
\centering
\includegraphics[width=7cm,angle=270]{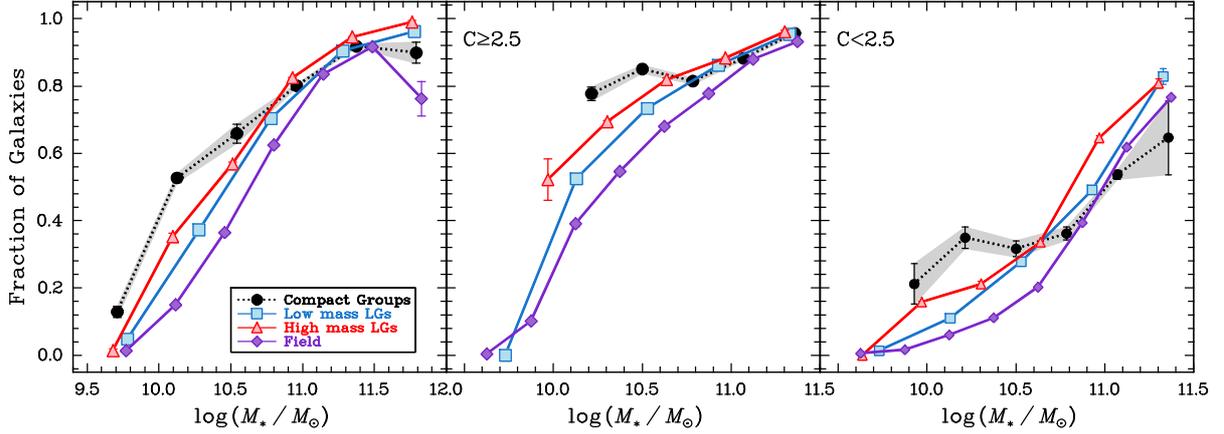}
\caption{Fraction of old stellar populations (D$_n(4000) \geq 1.6$) as a function of the stellar mass. Vertical
error-bars are obtained by using the bootstrap resampling technique. Filled black circles and dotted
lines represent the galaxies in CGs, low-mass LGs objects are shown as filled squares and solid blue
lines; high-mass LGs sample are shown as filled triangles and solid red lines. Galaxies in the field
are represented as filled diamonds and solid violet lines. \textit{Left panel} considers the whole 
sample, \textit{central panel} the early-type galaxies, while \textit{right panel} late-types. 
The binsizes in stellar mass are 0.5 dex for LGs, 0.4 dex for CGs and 0.3 dex for the field.
Abscissas are the medians of the mass within each bin. Bins with lesser than 10 galaxies were 
excluded.}
\label{fig:fd4000}
\end{figure*}
\subsection{The galaxy sample}
\label{subsec:galsample}
The MPA-JHU data contains the derived galaxy properties 
from the emission line analysis for the DR7 of the SDSS. The catalogue provides stellar masses based
on fits to the photometry following \citet{Kauffmann:2003} and \citet{Salim07}; star formation rates
based on \citet{Brinchmann:2004} and gas phase metallicity following \citet{Tremonti:2004}. The 
MPA-JHU data is based on a re-analysis of the spectra in order to take more care in the extraction 
of emission line fluxes than is done in the general purpose SDSS pipeline. The MPA-JHU also provides
the index D$_n(4000)$ based on the deﬁnition from \citet{Balogh98}. The D$_n(4000)$ index is a good 
indicator of the stellar population age. For composite galaxies (up to 40 per cent of their 
$H_\alpha$ luminosity come from AGN), the MPA-JHU data use the D$_n(4000)$ to estimate the SSFR. 
The parameters considered in our analysis are the stellar mass, the D$_n(4000)$ index and the SSFR. 
Average uncertainties in these quantities are: 0.13 dex, 0.5 and 0.5 dex, respectively.

As in Paper I, we compare the properties of galaxies in CGs and LGs with the properties of field 
galaxies. We consider as field galaxies to all DR7 MGS galaxies that were not identified as 
belonging to LGs by ZM11 groups or to CGs by \citet{McConnachie:2009}, and have apparent magnitudes
$14.5\le r\le 17.77$. For an adequate comparison with our samples of galaxies in groups, we use the 
same Monte Carlo algorithm of the previous subsection to construct a sample of field galaxies that 
has a similar redshift distribution as that of galaxies in our CG sample. This field sample includes
200102 galaxies.
\section{Comparing properties of galaxies: stellar mass, Specific Star Formation Rate and 
D$_n(4000)$}
\label{sec:comparison}
Analogously to Paper I, we classify galaxies into early and late types according to their 
concentration index. Typically, early-type galaxies have $C\geq2.5$, while for late-types $C<2.5$ 
\citep{Strateva:2001}. 
In this paper we compare properties of galaxies in samples drawn from an apparent magnitude
limited catalogue. Given a quantity bin, galaxies in it will have a range of absolute magnitudes 
and therefore a range of volume over which they could be detected in SDSS. In order to compensate 
for this, hereinafter we weight galaxy properties by using the  $1/V_{max}$ method 
\citep{Schmidt:68}, determining the maximum volume $V_{max}$ for which each galaxy could have 
been found.

We compare the normalised distributions of the stellar mass of galaxies in CGs and LGs and no 
differences are found, while field galaxies are lightly less massive than galaxies in other 
environments. We do not attempt to analyse the distributions of the metallicity because only 
$\sim 12$\% of galaxies in CGs have measured metallicity.
\begin{figure}[t]
\centering
\includegraphics[width=9cm]{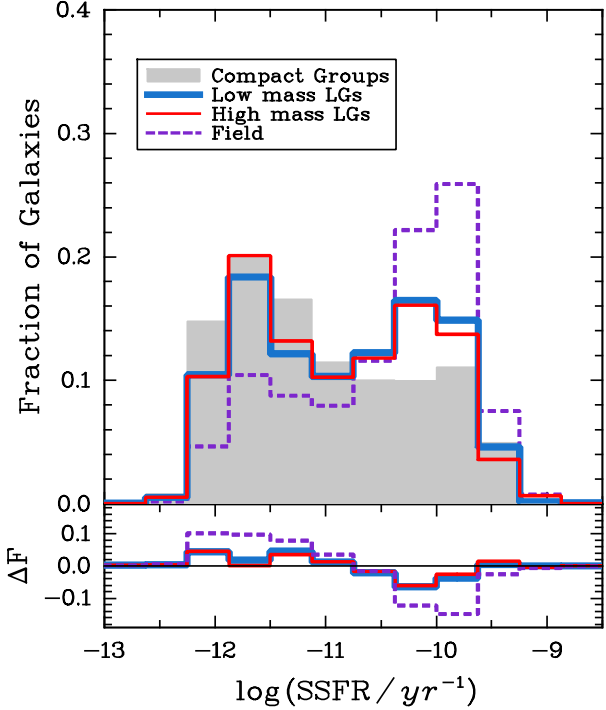}
\caption{Distributions of the SSFR in our samples. All distributions have been normalised to have 
the same area. Below each panel we show the residuals between the CG's and the other distributions. 
KS tests confirm that, there is no pair of them drawn from the same underlying distribution. 
The binsize is 0.37.
Colours and line types as in Fig. \ref{fig:d4000_tot}.} 
\label{fig:ssfr_tot}
\end{figure}
\begin{figure}[thb]
\centering
\includegraphics[width=9cm]{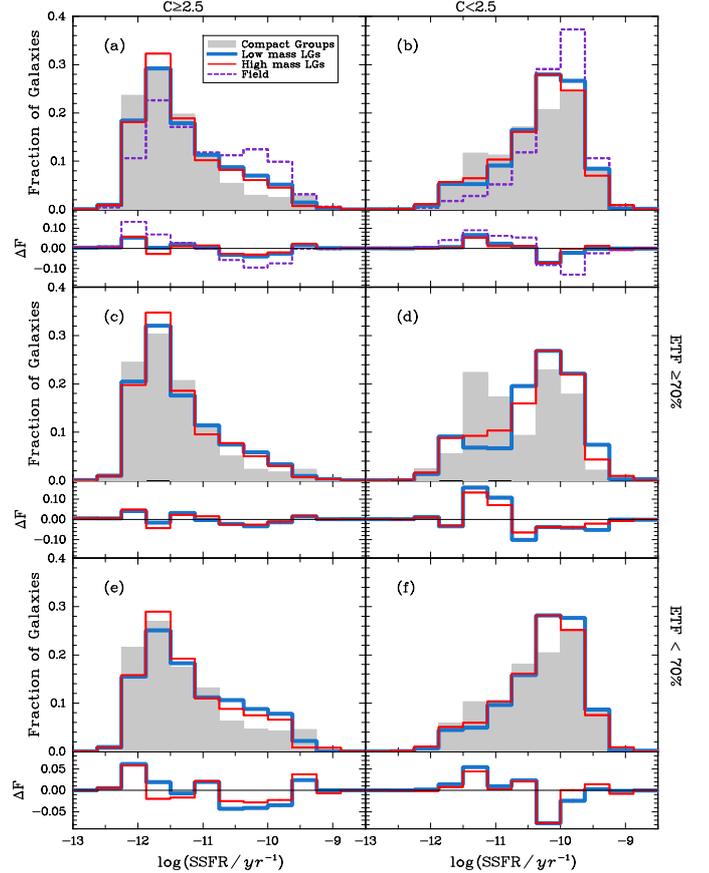}
\caption{Distributions of the specific star formation rate (SSFR) for early and late-type galaxies, 
{\em panels} as in Fig. \ref{fig:d4000}. All distributions have been normalised to have the same 
area. Below each panel we show as the residuals between the CG's and the other distributions. 
The binsize is 0.37.
KS tests confirm that, there is no pair of them drawn from the same underlying distribution.
Colours and line types as in Fig. \ref{fig:d4000_tot}.} 
\label{fig:ssfr}
\end{figure}
\subsection{Environment vs. D$_n(4000)$}
\label{sec:d4000}
Fig. \ref{fig:d4000_tot} compares the normalised distributions of the D$_n(4000)$ index for galaxies
in CGs with LG and field galaxies. Below each panel of Fig. \ref{fig:d4000_tot} we show the 
residuals between the distributions, i.e., for the property $X$, the difference 
$\Delta F(X)=f_{CG}(X)-f(X)$, where $f_{CG}(X)$ and $f(X)$ are the fractions of galaxies in the bin
centred on $X$ in the CGs and in the other sample, respectively. Kolmogorov-Smirnov tests (KS) 
confirm that, among the distributions shown, there is no pair of them drawn from the same underlying
distribution. As can be seen from Fig. \ref{fig:d4000_tot}, 
the stellar populations in galaxies in CGs tend to be older than in field or LG galaxies.
We find no difference between low or high mass LGs. CGs contain an excess of 
galaxies with D$_n(4000)\gtrsim 1.75$ and a deficit with D$_n(4000)\lesssim 1.75$. 
The stellar populations in field galaxies are the youngest. Our results agree with the 
comparison of CGs and field galaxies by \citet{Proctor:2004}, \citet{MdO:2005} and 
\citet{delaRosa:2007}.

We compare in Fig. \ref{fig:d4000} the normalised distributions of the D$_n(4000)$ index of late 
(\textit{right panels}) and early-type galaxies (\textit{left panels}). In this figure, we also 
analyse separately groups with a large ($\geq70\%$) fraction of early-type galaxies (ETF) 
({\textit{central panels}), and groups with a lower ETF (\textit{bottom panels}). As pointed out by
\citet{Bitsakis:2010}, if a group is dominated by early-type galaxies, it is more likely to be 
dynamically old, because interactions and mergers have had to occur in order to produce those 
galaxies. Conversely, groups not dominated by early-types, could be considered as dynamically young 
since some of their galaxies may have been gravitationally interacting for the first time. Thus, we 
use the ETF as an indication of group age. In the computation of the ETF we use all galaxies 
available in the CG catalogue, not only those that have spectroscopic data. In all panels of Fig. 
\ref{fig:d4000}, it is clear that both, early and late-type galaxies in CGs tend to have 
older stellar populations than in the other environments. The most significant result from this 
figure is that when we consider groups with large ETF: on the one hand, the stellar populations in 
early-type galaxies in both CGs and LGs have similar ages (\textit{panel} (c)), on the other hand, the stellar populations in late-type galaxies in CGs are on average much older than in
LGs (\textit{panel} (d)). This is in agreement with previous findings as 
\citet{Bitsakis:2010,Bitsakis:2011}.

Given that galaxy properties strongly depend on stellar mass \citep{Kauff:2003}, we explore the 
dependence of the D$_n(4000)$ on stellar mass. Fig. \ref{fig:fd4000} shows in its 
\textit{left panel} the fraction of galaxies with D$_n(4000)\geq 1.6$ as a function of stellar mass,
and in its \textit{central} and \textit{right panels}, the corresponding fractions for early and 
late-type galaxies, respectively. We use the criteria of \citet{Tinker:2011} to separate
between galaxies with old and young stellar populations. In particular, galaxies with D$_n(4000)\geq 1.6$ have old stellar populations.
This criteria has been used by other authors (\citealt{Geha:2012}, \citealt{Krause:2013}).
Over the whole range of stellar mass the fraction of galaxies with
old stellar populations increases with mass and is larger moving from the field to massive LGs.
Galaxies in CGs, however, show differences. For stellar masses below 
$\log(M_{\ast}/M_{\odot})\sim 10.5$, their stellar populations are the oldest irrespective of their type. On the other 
hand, for higher stellar masses, the CG environment can not be distinguished from LGs whether we 
consider all galaxies or just the early-types. The fraction of late-types with old stellar 
populations is consistent with field values, or it is marginally smaller, for masses
$\log(M_{\ast}/M_{\odot})\gtrsim 10.8$. This could be an indication that an important fraction of 
massive late-type galaxies in CGs have undergone important episodes of relatively recent star 
formation.

\subsection{Environment vs. SSFR}
\label{sec:ssfr}
Fig. \ref{fig:ssfr_tot} shows the distributions of the SSFR for our different samples of galaxies.
SSFR values show a clear bimodality, more markedly on LGs, but still present in the field and in 
CGs. Clearly, the population of field galaxies is dominated by galaxies with 
$\log({\rm SSFR/yr^{-1}})>-11$, and the opposite is seen in CGs. LGs show an intermediate behaviour 
and a more clear bimodality (e.g. \citealt{Wetzel:2011}).

In Fig. \ref{fig:ssfr} we split the galaxy samples into early and late-types and also distinguish 
whether the galaxies inhabit groups with ETF$\geq70\%$ or lesser. As expected, the bimodality seen 
in Fig. \ref{fig:ssfr_tot} can be explained in terms of early and late-type galaxies. It is also 
clear that both, early and late-types in CGs have lower values of SSFR compared to LG and even 
lower when compared to field galaxies. Of particular interest is \textit{panel} (d): here we find a 
marked bimodality in the SSFR of late-type galaxies in CGs with ETF$\geq70\%$, i.e., two populations
of late-type galaxies similar in numbers, one with 'normal' star formation and another with quenched
star formation, both of them inhabiting old CGs. A bimodality in the distribution of the SSFR in 
CGs galaxies has been reported previously by \citet{Johnson:2007}, \cite{Tzanavaris:2010}, 
\cite{Walker:2012}, \citet{Bitsakis:2010,Bitsakis:2011}.
 
We further explore the SSFR of galaxies in Fig. \ref{fig:fsfr}, where we show the fraction of
galaxies with high SSFR ($\log({\rm SSFR/yr^{-1}})\ge-11$) as a function of stellar mass. Field and 
LG galaxies have smooth, featureless, decreasing trends with stellar mass, over the whole mass 
range. At fixed mass, the fraction of star forming galaxies is larger in the field, followed by low
mass LGs and high mass LGs. Regarding CGs, for masses below $\log(M_{\ast}/M_{\odot})\sim 10.5$, the
fraction of star forming galaxies is lower than in the other environments (\textit{left panel}).
This is also true for late-type galaxies (\textit{right panel}). For higher masses, CGs are not 
distinguishable from LGs, considering the whole population or just the early-types 
(\textit{central panel}). This is not the case, however, of the fraction of star forming late-types.
It is consistent with, though even marginally higher than, the field value. All trends observed in 
this figure are in agreement with Fig. \ref{fig:fd4000}. 

In the light of Figs. \ref{fig:ssfr} (\textit{panel} (d)) and \ref{fig:fsfr}, we further explore the
SSFR of late-type galaxies in Fig. \ref{fig:ssfr_late}. In this figure we split galaxies not only 
as whether they are in large ETF groups but also distinguishing galaxies with 
$\log(M_{\ast}/M_{\odot})\ge10.5$ or lower. The strong gap in the SSFR is seen only in CGs, it is 
clearly present in galaxies inhabiting large ETF groups for both stellar mass ranges. The gap is 
particularly strong for massive late type galaxies. This further suggests that the high-density, 
low-velocity dispersion environment of CGs has accelerated the transition of galaxies from star 
forming to quiescent. Similar results have been reported by \citet{Tzanavaris:2010} and 
\citet{Plauchu:2012}.
\begin{figure*}[th]
\centering
\includegraphics[width=7cm,angle=270]{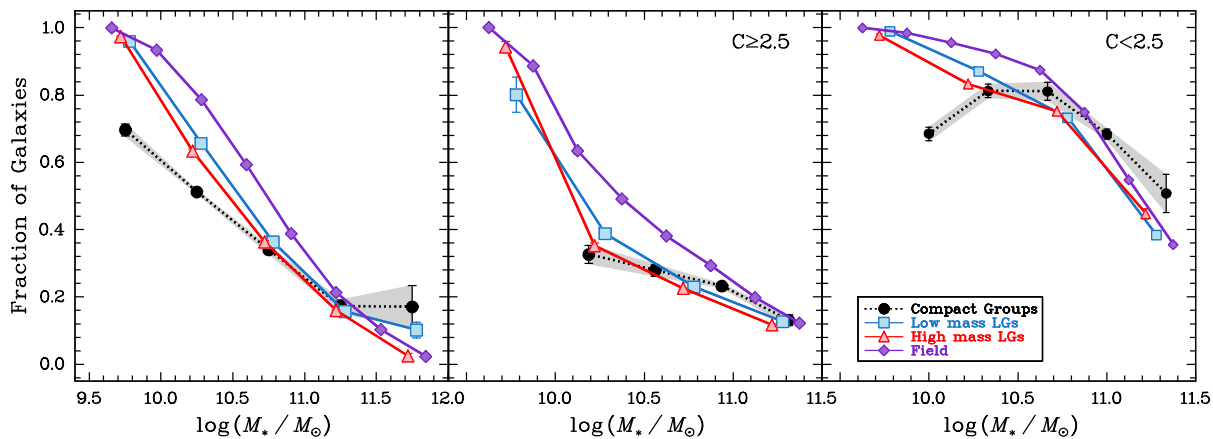}
\caption{Fraction of star forming galaxies according to their star formation rate 
($\log$(SSFR)$\ge-11$) as a function of the stellar mass. Vertical error-bars are obtained by using
the bootstrap resampling technique. \textit{Left panel} considers the whole samples of galaxies, 
\textit{central panel} the early-type galaxies, while \textit{right panel} late-types. 
The binsizes in stellar mass are 0.5 dex for LGs, 0.4 dex for CGs and 0.3 dex for the field.
Abscissas are the medians of the mass within each bin. Bins with lesser than 10 galaxies were 
excluded. Symbols and lines types as in Fig. \ref{fig:fd4000}.} 
\label{fig:fsfr}
\end{figure*}
\section{Conclusions and Discussion}
\label{sec:discussion}
We study the specific star formation rate and ages of galaxies in compact groups, loose groups and 
in the field in the redshift range $0.06<z<0.18$. We select samples of galaxies in CGs drawn from 
\citet{McConnachie:2009}, and in LGs taken from ZM11. We construct two samples of LG: low 
($\log(\Mvir/M_{\odot}h^{-1})\le 13.2$) and high ($\log(\Mvir/M_{\odot}h^{-1})\ge 13.6$) virial 
mass, both samples bound to have similar redshift distribution as the CG sample. Similarly, our 
sample of field galaxies was drawn to reproduce the redshift distribution of CGs. Galaxy properties
used in our work were taken from the MGS sample of the SDSS DR7 and the MPA-JHU DR7 release. The 
final samples have 748, 2319, 2352 of compact, low-mass, high-mass loose groups respectively. The 
corresponding number of member galaxies are: 978, 6996 and 7055. The field sample comprises 
200102 galaxies. 

We classify galaxies into early and late-types according to their concentration index. Following 
\cite{Bitsakis:2011}, we distinguish groups of galaxies as dynamically old or dynamically young. A 
group is classified as dynamically old if more than $70\%$ of its galaxies are early-types, or as 
dynamically young if otherwise. 
\begin{figure}[tbh]
\centering
\includegraphics[width=9cm]{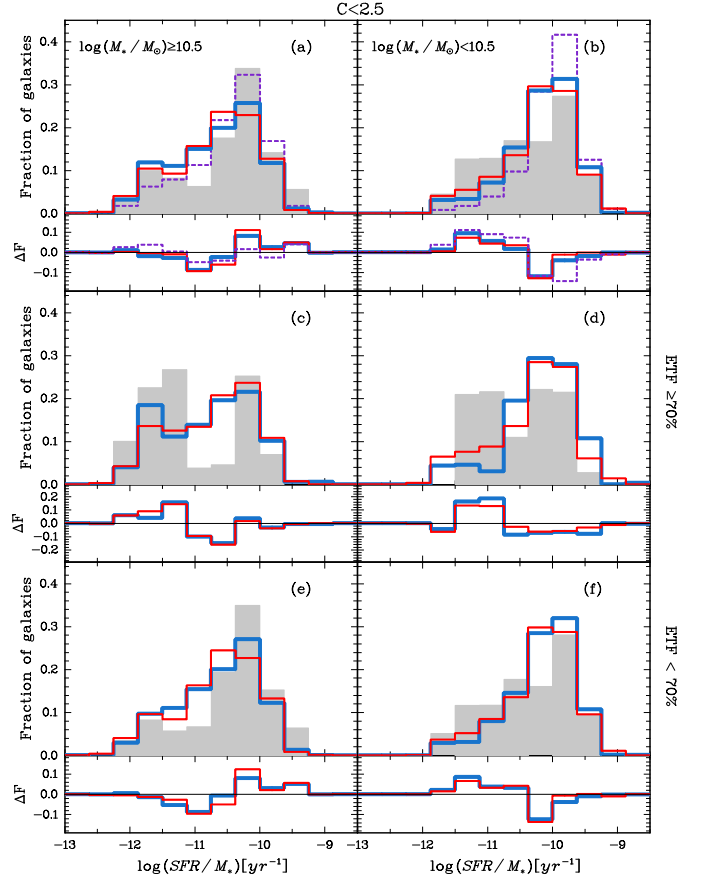}
\caption{Distributions of the specific star formation rate (SSFR) of late-type galaxies for 
$\log(M_{\ast}/M_{\odot})\ge10.5$ (\textit{left panels}) and for $\log(M_{\ast}/M_{\odot})<10.5$ 
(\textit{right panels}). All distributions have been normalised to have the same area. Below each 
panel we show the residuals between the CG's and the other distributions. 
The binsize is 0.37. KS tests confirm that, 
there is no pair of them drawn from the same underlying distribution. However, the distributions 
between low and high mass LG samples in \textit{panels} (a) and (c), are the same above 50\%.
Colours and line types as in Fig. \ref{fig:d4000_tot}.} 
\label{fig:ssfr_late}
\end{figure}
We find that the stellar populations in galaxies in CGs are, on average, older than in LGs or in the field. This agrees with 
\citet{Proctor:2004}, \citet{MdO:2005}, \citet{delaRosa:2007} and \cite{Plauchu:2012}. The stellar populations in late-type 
galaxies in CGs are on average, much older than in LGs. This is in agreement with 
\citet{Bitsakis:2010,Bitsakis:2011}. For stellar masses $\log(M_{\ast}/M_{\odot})\lesssim 10.5$, 
both, the stellar populations of early and late-types in CGs are older than in the other 
environments. For higher stellar masses, in early-types in CGs the stellar populations have ages 
similar to their LGs counterparts, and in late-types in CGs are of similar age to field galaxies, 
or even younger.

The distribution of SSFR is clearly bimodal, more markedly in LGs but still present in CGs and in 
the field. This bimodality has been reported in LGs by \citet{Wetzel:2011} and by 
\citet{Johnson:2007}, \citet{Tzanavaris:2010}, \citet{Bitsakis:2010,Bitsakis:2011} and
\citet{Walker:2012} in CGs. The bimodal distribution can be explained in terms of two galaxy 
populations: early and late-types.

Early and late-type galaxies in CGs have, on average, lower SSFR values than in LGs and the field. 
For $\log(M_{\ast}/M_{\odot})\lesssim 10.5$, CGs have the lowest fraction of star forming galaxies 
irrespective of galaxy type. For higher masses, the fraction of star forming early-types in CGs is 
comparable to LGs, and the fraction of star forming late-types in CGs is consistent with, or even 
higher than, the field. The distribution of SSFR for late-type galaxies in CGs shows a bimodality, 
with a strong gap for groups dynamically old. No such gap is found in any of the other environments.

Our findings suggest that, compared to the other environments, early-type galaxies in CGs have 
formed their stars and depleted their gas content more rapidly. This is an indication that these 
galaxies have had more frequent mergers and multiple past interactions, which is something expected
within an environment characterised by its high density and its low velocity dispersion.

For late-type galaxies in dynamically old CGs, we have found evidence of two populations of galaxies
regarding their SSFR. One population of late-types have 'normal' SSFR and they are forming stars, 
while the other population shows lower values of SSFR. Star forming late-types in CGs may be recent
acquisitions. By falling into CGs, these galaxies increase their star formation and rapidly consume
their gas. Further gas depletion and subsequent star formation quenching may be accounted for 
through merger events.

It is well known that groups of galaxies play a central role in accelerating galaxy evolution by 
enhancing star formation process and leading a transition to quiescence. The results of this series 
of papers, point out to the fact that compact groups are clearly more efficient than loose groups in
transforming galaxies. It is clear that the unique characteristics of compact groups, namely, their
high spatial density and low velocity dispersion, makes them one of the most extreme environments 
for galaxies, an environment where the transition from star forming to quiescence takes place in 
particularly short time scales.     
\begin{acknowledgements}
This work has been supported with grants from CONICET (PIP 11220080102603 and 11220100100336) and 
SECYT-UNC, Argentina. Funding for the Sloan Digital Sky Survey (SDSS) has been provided by the 
Alfred P. Sloan Foundation, the Participating Institutions, the National Aeronautics and Space 
Administration, the National Science Foundation, the U.S. Department of Energy, the Japanese 
Monbukagakusho, and the Max Planck Society. The SDSS Web site is http://www.sdss.org/.
The SDSS is managed by the Astrophysical Research Consortium (ARC) for the Participating 
Institutions. The Participating Institutions are The University of Chicago, Fermilab, the Institute 
for Advanced Study, the Japan Participation Group, The Johns Hopkins University, the Korean 
Scientist Group, Los Alamos National Laboratory, the Max Planck Institut f\"ur Astronomie (MPIA), 
the Max Planck Institut f\"ur Astrophysik (MPA), New Mexico State University, University of 
Pittsburgh, University of Portsmouth, Princeton University, the United States Naval Observatory, 
and the University of Washington.
\end{acknowledgements}
\bibliographystyle{aa} 

\begin{thebibliography}{45}
\expandafter\ifx\csname natexlab\endcsname\relax\def\natexlab#1{#1}\fi

\bibitem[{{Abazajian} {et~al.}(2009){Abazajian}, {Adelman-McCarthy},
  {Ag{\"u}eros}, {Allam}, {Allende Prieto}, {An}, {Anderson}, {Anderson},
  {Annis}, {Bahcall}, \& et~al.}]{dr7}
{Abazajian}, K.~N., {Adelman-McCarthy}, J.~K., {Ag{\"u}eros}, M.~A., {et~al.}
  2009, \apjs, 182, 543

\bibitem[{{Adelman-McCarthy} {et~al.}(2008){Adelman-McCarthy}, {Ag{\"u}eros},
  {Allam}, {Allende Prieto}, {Anderson}, {Anderson}, {Annis}, {Bahcall},
  {Bailer-Jones}, {Baldry}, {Barentine}, \& et~al.}]{dr6}
{Adelman-McCarthy}, J.~K., {Ag{\"u}eros}, M.~A., {Allam}, S.~S., {et~al.} 2008,
  \apjs, 175, 297

\bibitem[{{Bai} {et~al.}(2010){Bai}, {Rasmussen}, {Mulchaey}, {Dariush},
  {Raychaudhury}, \& {Ponman}}]{Bai:2010}
{Bai}, L., {Rasmussen}, J., {Mulchaey}, J.~S., {et~al.} 2010, \apj, 713, 637

\bibitem[{{Balogh} {et~al.}(1998){Balogh}, {Schade}, {Morris}, {Yee},
  {Carlberg}, \& {Ellingson}}]{Balogh98}
{Balogh}, M.~L., {Schade}, D., {Morris}, S.~L., {et~al.} 1998, \apjl, 504, L75

\bibitem[{{Beers} {et~al.}(1990){Beers}, {Flynn}, \& {Gebhardt}}]{Beers:1990}
{Beers}, T.~C., {Flynn}, K., \& {Gebhardt}, K. 1990, \aj, 100, 32

\bibitem[{{Bitsakis} {et~al.}(2011){Bitsakis}, {Charmandaris}, {da Cunha},
  {D{\'{\i}}az-Santos}, {Le Floc'h}, \& {Magdis}}]{Bitsakis:2011}
{Bitsakis}, T., {Charmandaris}, V., {da Cunha}, E., {et~al.} 2011, \aap, 533,
  A142

\bibitem[{{Bitsakis} {et~al.}(2010){Bitsakis}, {Charmandaris}, {Le Floc'h},
  {D{\'{\i}}az-Santos}, {Slater}, {Xilouris}, \& {Haynes}}]{Bitsakis:2010}
{Bitsakis}, T., {Charmandaris}, V., {Le Floc'h}, E., {et~al.} 2010, \aap, 517,
  A75

\bibitem[{{Blanton} {et~al.}(2003){Blanton}, {Brinkmann}, {Csabai}, {Doi},
  {Eisenstein}, {Fukugita}, {Gunn}, {Hogg}, \& {Schlegel}}]{Blanton:2003}
{Blanton}, M.~R., {Brinkmann}, J., {Csabai}, I., {et~al.} 2003, \aj, 125, 2348

\bibitem[{{Brinchmann} {et~al.}(2004){Brinchmann}, {Charlot}, {White},
  {Tremonti}, {Kauffmann}, {Heckman}, \& {Brinkmann}}]{Brinchmann:2004}
{Brinchmann}, J., {Charlot}, S., {White}, S.~D.~M., {et~al.} 2004, \mnras, 351,
  1151

\bibitem[{{Coenda} {et~al.}(2012){Coenda}, {Muriel}, \&
  {Mart{\'{\i}}nez}}]{Coenda:2012}
{Coenda}, V., {Muriel}, H., \& {Mart{\'{\i}}nez}, H.~J. 2012, \aap, 543, A119

\bibitem[{{de la Rosa} {et~al.}(2007){de la Rosa}, {de Carvalho}, {Vazdekis},
  \& {Barbuy}}]{delaRosa:2007}
{de la Rosa}, I.~G., {de Carvalho}, R.~R., {Vazdekis}, A., \& {Barbuy}, B.
  2007, \aj, 133, 330

\bibitem[{{Dressler}(1980)}]{Dressler:1980}
{Dressler}, A. 1980, \apjs, 42, 565

\bibitem[{{Geha} {et~al.}(2012){Geha}, {Blanton}, {Yan}, \&
  {Tinker}}]{Geha:2012}
{Geha}, M., {Blanton}, M.~R., {Yan}, R., \& {Tinker}, J.~L. 2012, \apj, 757, 85

\bibitem[{{Hickson}(1982)}]{Hickson:1982}
{Hickson}, P. 1982, \apj, 255, 382

\bibitem[{{Hirschmann} {et~al.}(2014){Hirschmann}, {De Lucia}, {Wilman},
  {Weinmann}, {Iovino}, {Cucciati}, {Zibetti}, \&
  {Villalobos}}]{Hirschmann:2014}
{Hirschmann}, M., {De Lucia}, G., {Wilman}, D., {et~al.} 2014, ArXiv e-prints

\bibitem[{{Huchra} \& {Geller}(1982)}]{H&G:1982}
{Huchra}, J.~P. \& {Geller}, M.~J. 1982, \apj, 257, 423

\bibitem[{{Johnson} {et~al.}(2007){Johnson}, {Hibbard}, {Gallagher},
  {Charlton}, {Hornschemeier}, {Jarrett}, \& {Reines}}]{Johnson:2007}
{Johnson}, K.~E., {Hibbard}, J.~E., {Gallagher}, S.~C., {et~al.} 2007, \aj,
  134, 1522

\bibitem[{{Kauffmann} {et~al.}(2003{\natexlab{a}}){Kauffmann}, {Heckman},
  {White}, {Charlot}, {Tremonti}, {Brinchmann}, {Bruzual}, {Peng}, {Seibert},
  {Bernardi}, {Blanton}, {Brinkmann}, {Castander}, {Cs{\'a}bai}, {Fukugita},
  {Ivezic}, {Munn}, {Nichol}, {Padmanabhan}, {Thakar}, {Weinberg}, \&
  {York}}]{Kauffmann:2003}
{Kauffmann}, G., {Heckman}, T.~M., {White}, S.~D.~M., {et~al.}
  2003{\natexlab{a}}, \mnras, 341, 33

\bibitem[{{Kauffmann} {et~al.}(2003{\natexlab{b}}){Kauffmann}, {Heckman},
  {White}, {Charlot}, {Tremonti}, {Peng}, {Seibert}, {Brinkmann}, {Nichol},
  {SubbaRao}, \& {York}}]{Kauff:2003}
{Kauffmann}, G., {Heckman}, T.~M., {White}, S.~D.~M., {et~al.}
  2003{\natexlab{b}}, \mnras, 341, 54

\bibitem[{{Krause} {et~al.}(2013){Krause}, {Hirata}, {Martin}, {Neill}, \&
  {Wyder}}]{Krause:2013}
{Krause}, E., {Hirata}, C.~M., {Martin}, C., {Neill}, J.~D., \& {Wyder}, T.~K.
  2013, \mnras, 428, 2548

\bibitem[{{Limber} \& {Mathews}(1960)}]{Limber:1960}
{Limber}, D.~N. \& {Mathews}, W.~G. 1960, \apj, 132, 286

\bibitem[{{Mart{\'{\i}}nez} {et~al.}(2013){Mart{\'{\i}}nez}, {Coenda}, \&
  {Muriel}}]{Martinez:2013}
{Mart{\'{\i}}nez}, H.~J., {Coenda}, V., \& {Muriel}, H. 2013, \aap, 557, A61

\bibitem[{{Mart{\'{\i}}nez} {et~al.}(2002){Mart{\'{\i}}nez}, {Zandivarez},
  {Dom{\'{\i}}nguez}, {Merch{\'a}n}, \& {Lambas}}]{MZDML:2002}
{Mart{\'{\i}}nez}, H.~J., {Zandivarez}, A., {Dom{\'{\i}}nguez}, M.,
  {Merch{\'a}n}, M.~E., \& {Lambas}, D.~G. 2002, \mnras, 333, L31

\bibitem[{{McConnachie} {et~al.}(2009){McConnachie}, {Patton}, {Ellison}, \&
  {Simard}}]{McConnachie:2009}
{McConnachie}, A.~W., {Patton}, D.~R., {Ellison}, S.~L., \& {Simard}, L. 2009,
  \mnras, 395, 255

\bibitem[{{McGee} {et~al.}(2011){McGee}, {Balogh}, {Wilman}, {Bower},
  {Mulchaey}, {Parker}, \& {Oemler}}]{McGee:2011}
{McGee}, S.~L., {Balogh}, M.~L., {Wilman}, D.~J., {et~al.} 2011, \mnras, 413,
  996

\bibitem[{{Mendes de Oliveira} {et~al.}(2005){Mendes de Oliveira}, {Coelho},
  {Gonz{\'a}lez}, \& {Barbuy}}]{MdO:2005}
{Mendes de Oliveira}, C., {Coelho}, P., {Gonz{\'a}lez}, J.~J., \& {Barbuy}, B.
  2005, \aj, 130, 55

\bibitem[{{Oemler}(1974)}]{Oemler:1974}
{Oemler}, A.~J. 1974, \apj, 194, 1

\bibitem[{{Petrosian}(1976)}]{petro76}
{Petrosian}, V. 1976, \apjl, 209, L1

\bibitem[{{Plauchu-Frayn} {et~al.}(2012){Plauchu-Frayn}, {Del Olmo}, {Coziol},
  \& {Torres-Papaqui}}]{Plauchu:2012}
{Plauchu-Frayn}, I., {Del Olmo}, A., {Coziol}, R., \& {Torres-Papaqui}, J.~P.
  2012, \aap, 546, A48

\bibitem[{{Proctor} {et~al.}(2004){Proctor}, {Forbes}, {Hau}, {Beasley}, {De
  Silva}, {Contreras}, \& {Terlevich}}]{Proctor:2004}
{Proctor}, R.~N., {Forbes}, D.~A., {Hau}, G.~K.~T., {et~al.} 2004, \mnras, 349,
  1381

\bibitem[{{Rasmussen} {et~al.}(2012){Rasmussen}, {Mulchaey}, {Bai}, {Ponman},
  {Raychaudhury}, \& {Dariush}}]{Rasmussen:2012}
{Rasmussen}, J., {Mulchaey}, J.~S., {Bai}, L., {et~al.} 2012, \apj, 757, 122

\bibitem[{{Salim} {et~al.}(2007){Salim}, {Rich}, {Charlot}, {Brinchmann},
  {Johnson}, {Schiminovich}, {Seibert}, {Mallery}, {Heckman}, {Forster},
  {Friedman}, {Martin}, {Morrissey}, {Neff}, {Small}, {Wyder}, {Bianchi},
  {Donas}, {Lee}, {Madore}, {Milliard}, {Szalay}, {Welsh}, \& {Yi}}]{Salim07}
{Salim}, S., {Rich}, R.~M., {Charlot}, S., {et~al.} 2007, \apjs, 173, 267

\bibitem[{{Schlegel} {et~al.}(1998){Schlegel}, {Finkbeiner}, \&
  {Davis}}]{sch98}
{Schlegel}, D.~J., {Finkbeiner}, D.~P., \& {Davis}, M. 1998, \apj, 500, 525

\bibitem[{{Schmidt}(1968)}]{Schmidt:68}
{Schmidt}, M. 1968, \apj, 151, 393

\bibitem[{{Scudder} {et~al.}(2012){Scudder}, {Ellison}, \&
  {Mendel}}]{Scudder:2012}
{Scudder}, J.~M., {Ellison}, S.~L., \& {Mendel}, J.~T. 2012, \mnras, 423, 2690

\bibitem[{{Strateva} {et~al.}(2001){Strateva}, {Ivezi{\'c}}, {Knapp},
  {Narayanan}, {Strauss}, {Gunn}, {Lupton}, {Schlegel}, {Bahcall}, \&
  et~al.}]{Strateva:2001}
{Strateva}, I., {Ivezi{\'c}}, {\v Z}., {Knapp}, G.~R., {et~al.} 2001, \aj, 122,
  1861

\bibitem[{{Strauss} {et~al.}(2002){Strauss}, {Weinberg}, {Lupton}, {Narayanan},
  {Annis}, {Bernardi}, {Blanton}, {Burles}, {Connolly}, {Dalcanton}, {Doi},
  {Eisenstein}, \& et~al.}]{Strauss:2002}
{Strauss}, M.~A., {Weinberg}, D.~H., {Lupton}, R.~H., {et~al.} 2002, \aj, 124,
  1810

\bibitem[{{Tinker} {et~al.}(2011){Tinker}, {Wetzel}, \& {Conroy}}]{Tinker:2011}
{Tinker}, J., {Wetzel}, A., \& {Conroy}, C. 2011, ArXiv e-prints

\bibitem[{{Tremonti} {et~al.}(2004){Tremonti}, {Heckman}, {Kauffmann},
  {Brinchmann}, {Charlot}, {White}, {Seibert}, {Peng}, {Schlegel}, {Uomoto},
  {Fukugita}, \& {Brinkmann}}]{Tremonti:2004}
{Tremonti}, C.~A., {Heckman}, T.~M., {Kauffmann}, G., {et~al.} 2004, \apj, 613,
  898

\bibitem[{{Tzanavaris} {et~al.}(2010){Tzanavaris}, {Hornschemeier},
  {Gallagher}, {Johnson}, {Gronwall}, {Immler}, {Reines}, {Hoversten}, \&
  {Charlton}}]{Tzanavaris:2010}
{Tzanavaris}, P., {Hornschemeier}, A.~E., {Gallagher}, S.~C., {et~al.} 2010,
  \apj, 716, 556

\bibitem[{{Walker} {et~al.}(2012){Walker}, {Johnson}, {Gallagher}, {Charlton},
  {Hornschemeier}, \& {Hibbard}}]{Walker:2012}
{Walker}, L.~M., {Johnson}, K.~E., {Gallagher}, S.~C., {et~al.} 2012, \aj, 143,
  69

\bibitem[{{Wetzel} {et~al.}(2011){Wetzel}, {Tinker}, \& {Conroy}}]{Wetzel:2011}
{Wetzel}, A.~R., {Tinker}, J.~L., \& {Conroy}, C. 2011, ArXiv e-prints

\bibitem[{{Wetzel} {et~al.}(2013){Wetzel}, {Tinker}, {Conroy}, \& {van den
  Bosch}}]{Wetzel:2013}
{Wetzel}, A.~R., {Tinker}, J.~L., {Conroy}, C., \& {van den Bosch}, F.~C. 2013,
  \mnras, 432, 336

\bibitem[{{York} {et~al.}(2000){York}, {Anderson}, {Anderson}, {Annis},
  {Bahcall}, \& et~al.}]{York:2000}
{York}, D.~G., {Anderson}, Jr., J.~E., {Anderson}, S.~F., {et~al.} 2000, \aj,
  120, 1579

\bibitem[{{Zandivarez} \& {Mart{\'{\i}}nez}(2011)}]{ZM11}
{Zandivarez}, A. \& {Mart{\'{\i}}nez}, H.~J. 2011, \mnras, 415, 2553

\end{thebibliography}

\end{document}